\documentclass
[%
aps,
reprint,
showpacs,
preprintnumbers,
amsmath,
amssymb,
pre,
]{revtex4-1}

\usepackage{graphicx}
\usepackage{float}
\usepackage{dcolumn}
\usepackage{bm}
\usepackage{color,soul}

\usepackage[utf8]{inputenc}
\usepackage[T1]{fontenc}
\usepackage{mathptmx}
\usepackage{etoolbox}

\begin{document}
\setulcolor{red}    
\setstcolor{blue}   
\sethlcolor{yellow} 


\title{Control of electron beam polarization in the bubble regime of laser-wakefield acceleration}

    \author{H. C. Fan$^{1}$}
	\author{X. Y. Liu$^{1}$}
	\author{X. F. Li$^{2,3,4}$}\email{xia.li@fz-juelich.de or xiaofengli@sjtu.edu.cn}%
	\author{J. F. Qu$^{1}$}
	\author{Q. Yu$^{5}$}
	\author{Q. Kong$^{1}$}\email{qkong@fudan.edu.cn}%
	\author{S. M. Weng$^{3,4}$}
	\author{M. Chen$^{3,4}$}
	\author{M. B\"uscher$^{6,7}$}
	\author{P. Gibbon$^{2,8}$}
	\author{S. Kawata$^{9}$}
	\author{Z. M. Sheng$^{3,4,10,11}$}

	\affiliation{$^1$ Key Laboratory of Nuclear Physics and Ion-beam Application (MOE), Institute of Modern Physics, Department of Nuclear Science and Technology, Fudan University, Shanghai 200433, China}
	\affiliation{$^2$ Institute for Advanced Simulation, J\"ulich Supercomputing Centre, Forschungszentrum J\"ulich, 52425 Jülich, Germany}%
	\affiliation{$^3$ Key Laboratory for Laser Plasmas (MoE), School of Physics and Astronomy,Shanghai Jiao Tong University, Shanghai 200240, China}%
    \affiliation{$^4$ Collaborative Innovation Center of IFSA, Shanghai Jiao Tong University, Shanghai 200240, China}
	\affiliation{$^5$ State Key Laboratory of High Field Laser Physics, Shanghai Institute of Optics and Fine Mechanics, Chinese Academy of Sciences, Shanghai 201800, China}
	\affiliation{$^6$ Peter Gr\"unberg Institut (PGI-6), Forschungszentrum J\"ulich, Wilhelm-Johnen-Str. 1, 52425 Jülich, Germany }
	\affiliation{$^7$ Institut f\"ur Laser- und Plasmaphysik, Heinrich-Heine-Universit\"at D\"usseldorf, Universit\"atsstr. 1, 40225 D\"usseldorf, Germany}
	\affiliation{$^8$ Centre for Mathematical Plasma Astrophysics, Katholieke Universiteit Leuven, 3000 Leuven, Belgium}
	\affiliation{$^9$ Graduate School of Engineering, Utsunomiya University, Utsunomiya 321-8585, Japan}
	\affiliation{$^{10}$ SUPA, Department of Physics, University of Strathclyde, Glasgow G4 0NG, UK}
	\affiliation{$^{11}$ Tsung-Dao Lee Institute, Shanghai Jiao Tong University, Shanghai 200240, China}

\date{\today}

\begin{abstract}
Electron beam polarization in the bubble regime of the interaction between a high-intensity laser and a longitudinally pre-polarized plasma is investigated by means of the Thomas-Bargmann-Michel-Telegdi equation. Using a test-particle model, the dependence of the  accelerated electron polarization on the  bubble geometry is analyzed in detail. Tracking the polarization dynamics of individual electrons reveals that although the spin direction changes during both the self-injection process and acceleration phase, the former has the biggest impact. For nearly spherical bubbles, 
the polarization of electron beam persists after capture and acceleration in the bubble. By contrast, for aspherical bubble shapes, the electron beam becomes rapidly depolarized, and the net polarization direction can even reverse in the case of a oblate spheroidal bubble. These  findings are confirmed  via particle-in-cell simulations. 
\end{abstract}

\maketitle

\section{\label{sec:level1}INTRODUCTION}
Laser wakefield acceleration (LWFA) has made remarkable progress since it was first proposed by Tajima and Dawson in 1979 \cite{tajima1979laser} and experimentally realized through the rapid advancement of laser technology via chirped-pulse amplification (CPA) \cite{mourou1992development}. Since then,  improved understanding of various schemes such as plasma beat wave acceleration \cite{rosenbluth1972excitation}, multiple laser pulses \cite{shadwick2009nonlinear,umstadter1994nonlinear}, and self-modulated laser wakefield acceleration \cite{leemans2001gamma} have contributed to a series of milestones. Of particular note is the generation of quasimonoenergetic electron beams in the bubble regime \cite{pukhov2002laser}, which triggered significant experimental progress and widespread interest \cite{geddes2004high,mangles2004monoenergetic,faure2004laser}. In recent years, applications of wakefield acceleration have been actively pursued, such as synchrotron radiation sources \cite{jaroszynski2006radiation,schlenvoigt2008compact} and polarized particle beams \cite{wen2019polarized,wu2019polarized,buscher2020generation}. 

Spin-polarized particle beams are widely used in nuclear and particle physics to study the interaction and structure of matter, and to test the standard model of particle physics \cite{rathmann2014search,moortgat2008polarized}. In particular, the structure of subatomic particles like protons or neutrons can be explored to get further insights into quantum chromodynamics \cite{burkardt2009spin} or to probe the nuclear spin structure \cite{ageev2005measurement}. Additionally, polarized particle beams are advantageous to achieve a deeper understanding of nuclear reactions \cite{glashausser1979nuclear}, to investigate symmetry violations, to measure quantum numbers of new particles \cite{rathmann2014search,jaffe2003open,adlarson2014evidence,baer2013international}, or to investigate molecular dynamics \cite{gay2009physics,mcdaniel1982applied}. In 2019, An \emph{et al.} proposed to map electromagnetic field structures of plasmas by using a spin-polarized relativistic electron beam \cite{an2019mapping}.
Recently, polarized multi-GeV proton beams produced by ultra-intense laser interactions were studied via simulations \cite{li2021polarized}. In contrast, a first polarization measurement of few-MeV laser accelerated protons reported a negative result, \emph{i.e.\/} no polarization build-up during the acceleration process \cite{raab2014}.
At present, the preparation of polarized electron beams mainly relies on spontaneous polarization in the magnetic fields of storage rings due to the emission of spin-flip synchrotron radiation, \emph{i.e.\/} the well-known Sokolov-Ternov effect \cite{mane2005spin,sokolov1967synchrotron}. This technique requires conventional particle accelerators that are typically very large in scale and budget \cite{mane2005spin}.

The acceleration of polarized electron beams by means of laser-driven acceleration promises to be cost-efficient and highly effective. Despite many advances mostly on the theoretical side, several principal issues need to be addressed, for example: (i) is it possible to alter the polarization of an initially unpolarized target through interaction with relativistic laser pulses \cite{guo2020stochasticity,chen2019polarized,seipt2019ultrafast,del2017spin,li2019ultrarelativistic}? or (ii) are the spins so inert during the short acceleration period that a pre-polarized target is  required \cite{wen2019polarized,wu2019polarized,wu2020spin,wen2017spin,wu2019wakefield}? Following the work by H\"utzen \emph{et al.} \cite{huetzen2019proton},  Wen \emph{et al.} \cite{wen2019polarized} have proposed to generate high-current polarized electron beams in the interaction of an ultra-intense laser pulse with a pre-polarized gas plasma, which is produced through photo-dissociation by a circularly polarized ultra-violet (UV) laser pulse \cite{sofikitis2018ultrahigh}. The work of Vieira \emph{et al.} showed that  spin is depolarized mainly  in the injection phase \cite{vieira2011polarized}.

Previous works show that the distribution of the electromagnetic fields is  affected by the accelerating bubble geometry \cite{sadighi2010potential,whcheng2010transverse,xfli2015general}. It is thus likely that the self-injection process can be affected by the bubble shape \cite{li2014dependence,zahra2020}. Moreover, the work of Qu \emph{et al.} \cite{qu2019terahertz} indicates that the frequency of THz radiation generated by the shell electrons also depends on the bubble shape. In this paper, the evolution of the electron beam polarization injected into various shapes of ellipsoidal bubbles is discussed in detail, and it is found that the polarization of electron beams can be controlled by adjusting the bubble geometry. The results of our analysis are  highly relevant to experimental implementations of polarized electron beams, such as those planned at the European EuPRAXIA facility \cite{eupraxia2020}.

\section{\label{sec:level2}TEST-PARTICLE MODEL}
By choosing appropriate laser and plasma parameters, different shaped wakefield bubbles can be achieved \cite{li2014dependence}. A series of 2.5D particle-in-cell (PIC) simulation with the code EPOCH \cite{arber2015contemporary} was carried out to analyze the bubble geometry. The laser propagates in the $x$-direction with linear polarization in the $y$-direction and a Gaussian envelope 
\begin{equation}
    E=E_0\frac{w_0}{w(x)}\exp\left(-\frac{y^2+z^2}{w^2(x)}\right)\exp\left(-\frac{(k x-\omega t)^2}{(0.5\tau)^2}\right)\cos(\varphi),  \label{gaussian}
\end{equation}
where, $w(x)=w_0[1+(x-x_0)^2/{z_R^2}]^{0.5}$, with laser waist $w_0=10\lambda$, pulse duration $\tau=21fs$, laser intensity $a_0=eE_0/m_e\omega c=20$ and wavelength $\lambda=800$ nm. The $x_0=30\lambda$ is the position of the laser waist, and $z_R={\pi}w_0^2/\lambda$ is the Rayleigh length. The vacuum length was $30\lambda$ and the laser beam was focused at the left edge of plasma. The simulation box is $140\lambda(x)\times100\lambda(y)$ with resolution $\mathrm{d}x=\lambda/32$ and $\mathrm{d}y=5\mathrm{d}x$. There were 16 pseudo-particles per cell.  We define the aspect ratio $\eta =R_{\perp}/R_{\|}$ to describe the shape of bubble, where $R_{\parallel}$ and $R_{\perp}$ are the longitudinal and transverse radii respectively. Different $R_{\parallel}$ and $R_{\perp}$ can be obtained by changing the plasma density for the same laser system \cite{li2014dependence}. Following this definition, $\eta<1$ indicates a prolate spheroid, $\eta=1$ indicates a sphere, and $\eta>1$ indicates an oblate spheroid. For an initial plasma density of $n_0=0.011n_c$, we find that $R_{\parallel}=14.38\lambda$ and $R_{\perp}=13.13 \lambda$. Consequently, the aspect ratio $\eta=0.91<1$, represents a prolate spheroidal bubble.

\begin{figure*}[t]
	\centering
	\includegraphics[width=0.95\textwidth]{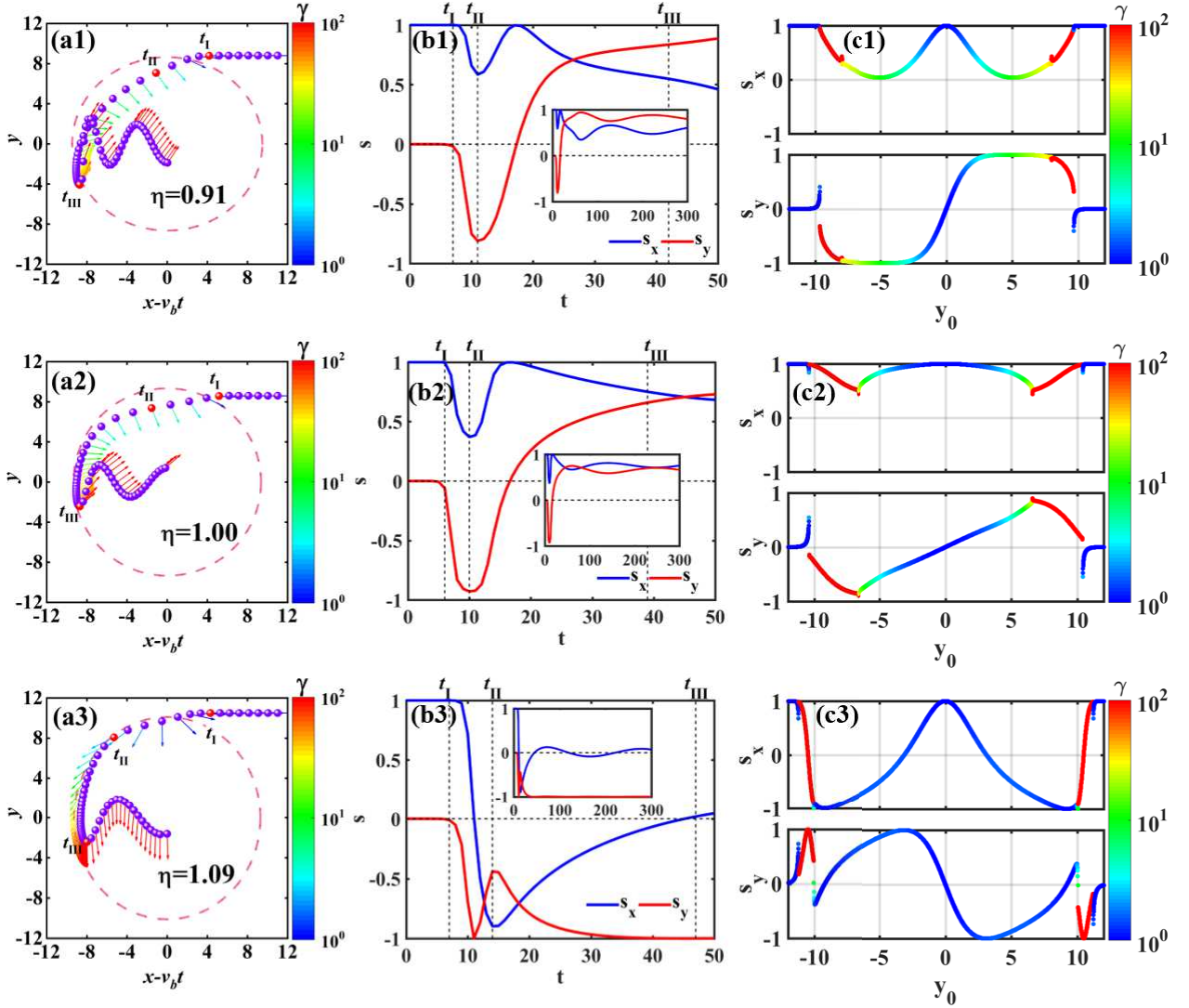}
	\caption{(color online) (a1, a2, a3) Trajectories (purple dots) and the spin (arrows) orientation of typical electrons in case 1 ($\eta=0.91$), case 2 ($\eta=1.00$) and case 3 ($\eta=1.09$), respectively. Initially, the electron located at the front of bubble and the spin is aligned with the propagation direction of laser. The colour of arrows indicates the electron energy. (b1, b2, b3) The electron trajectories of $s_x$ and $s_y$. (c1, c2, c3) The distribution of initial direction (${s}_x$) and perpendicular direction (${s}_y$) for electron spin as a function of initial position $y_0$ at the end of acceleration stage. The dot color indicates the electron energy, and the injected electrons ($\gamma>50$) are denoted as red dots. The radii of bubble ($R_{\perp}$ and $R_{\|}$) are obtained from PIC simulation, and the electron densities are $n_0=0.011n_c$, $n_0=0.0135n_c$ and $n_0=0.017n_c$ respectively. The laser spot size, duration and intensity are: $w_0=10\lambda$, $\tau=21fs$ and $a_0=20$.}
	\label{fig_1}
\end{figure*}

 The electromagnetic field distribution in a spherical bubble has already been theoretically and numerically studied \cite{kostyukov2004phenomenological,lu2006nonlinear,lu2007generating}. Based on the work of Li \emph{et al.} \cite{xfli2015general}, the electromagnetic field of an ellipsoidal bubble can be written as
\begin{subequations}
\label{eq:bubble field}
\begin{equation}
    E_x=\frac{\eta^2}{\eta^2(1-v_\mathrm{b}^2)+2} \xi, \label{eq:Ex}  \vspace{1ex}
\end{equation}
\begin{equation}
    E_y=\frac{2-\eta^2 v_\mathrm{b}^2}{2 \eta^2(1-v_\mathrm{b}^2)+4}y,\label{eq:Ey}
\end{equation}
\begin{equation}
    E_z=\frac{2-\eta^2 v_\mathrm{b}^2}{2\eta^2(1-v_\mathrm{b}^2)+4} z,\label{eq:Ez}
\end{equation}
\begin{equation}
    B_x=0, \label{eq:Bx}
\end{equation}
\begin{equation}
    B_y=\frac{v_\mathrm{b} \eta^2}{2 \eta^2(1-v_\mathrm{b}^2)+4} z, \label{eq:By}
\end{equation}
\begin{equation}
    B_z=-\frac{v_\mathrm{b} \eta^2}{2 \eta^2(1-v_\mathrm{b}^2)+4} y, \label{eq:Bz}
\end{equation}
\end{subequations}
where $\xi=x-v_bt$, $v_\mathrm{b}=\sqrt{1-\gamma_\mathrm{b}^{-2}}$  is the bubble phase velocity and  $\gamma_\mathrm{b}=0.45\sqrt{n_c/n_0}$  \cite{kostyukov2004phenomenological,schnell2013optical}. To confine the field distribution inside the bubble, a modified factor $f(r)=\left[\tanh({R_{\|}/d-r/d})+1\right]/2$ was used, $r=\sqrt{\xi^2+(y^2+z^2)/\eta^2)}$ and $d$ is the width of the electron sheath. In this work, $d=0.5$ was used. Here, we used dimensionless units, by normalizing the length to $k_p$, the velocity to $c$, the electron density to $n_0$, the electric field to $m_ec\omega_p/e$, and the magnetic field to $m_e\omega_p/e$.

Initially, the electron is at rest in front of the bubble with a position ($x_0$, $y_0$). Considering a fully polarized plasma, the electron spin at initial time is aligned with in the bubble propagation direction ($x$). To follow the trajectory of an electron in the bubble, a fourth-order Runge-Kutta method was adopted to numerically solve the relativistic Newton-Lorentz equation, ${\mathrm{d} \bm{P}}/{\mathrm{d} t}=-e\left[\bm{E}+{(\bm{P}}/{\gamma}) \times \bm{B}\right]$, where $\gamma=\sqrt{1-{v}^2/c^2}$ is the relativistic factor. Meanwhile, the spin precession of an electron in the electromagnetic field  was calculated according to the Thomas-Bargmann-Michel-Telegdi (TBMT) equation \cite{mane2005spin} $\mathrm{d} \bm{s} / \mathrm{d} t=\left(\bm{\Omega}_{T}+\bm{\Omega}_{a}\right) \times \bm{s}$ with
\begin{subequations}
\begin{equation}
    \bm{\Omega}_T=\frac{e}{m_e}\left(\frac{1}{\gamma}\bm{B}-\frac{1}{\gamma+1} \frac{\bm{v}}{c^2} \times \bm{E} \right),   \label{eq:omegaT}
\end{equation}
\begin{equation}
    \bm{\Omega}_a=\bm{a}_e \frac{e}{m_e} \left( \bm{B}-\frac{\gamma}{\gamma+1} \frac{\bm{v}}{c^2} (\bm{v}\cdot \bm{B})-\frac{\bm{v}}{c^2} \times \bm{E} \right),   \label{eq:omegaa}
\end{equation}
\end{subequations}
where $a_e\approx 1.16 \times 10^{-3}$ is the anomalous magnetic moment of the electron and the spin is normalized to $|\bm{s}|=1$. The Boris-rotation method was adopted to numerically solve the TBMT equation \cite{thomas2020boris}.

Owing to the azimuthal symmetry of the wake magnetic field, the electron orbit and spin can be placed in the same plane \cite{kostyukov2004phenomenological,wen2019polarized}, so for simplicity the $XY$ plane was adopted for this calculation. The simulation results of three typical cases are displayed in Figure 1. The aspect ratio of these bubbles  were  $\eta=0.91$ in case 1,  $\eta=1.00$ in case 2 and $\eta=1.09$ in case 3, representing prolate spheriodal, spherical and oblate spheroidal bubbles, respectively. The trajectory and the spin orientation of a typical electron are presented in  Fig. 1(a1)--(a3). The motion of electrons can be divided into four stages: 
(i) $t<t_{\mathrm{\uppercase\expandafter{\romannumeral1}}}$, the electron does not feel the bubble field; 
(ii) $t_{\mathrm{\uppercase\expandafter{\romannumeral1}}}<t<t_{\mathrm{\uppercase\expandafter{\romannumeral2}}}$, the electron is located on the bubble shell and its spin rotates clockwise, such that $s_y$ decreases from 0 to almost $-1$ as shown in Fig. 1(b1); 
(iii) $t_{\mathrm{\uppercase\expandafter{\romannumeral2}}}<t<t_{\mathrm{\uppercase\expandafter{\romannumeral3}}}$, the electron reaches  the tail of the bubble and its spin rotates counter-clockwise, and $s_y$ increases as revealed in Fig. 1(b1);
(iv) $t>t_{\mathrm{\uppercase\expandafter{\romannumeral3}}}$, the electron is captured in the bubble and its spin procession slows down.

\begin{figure*}[t]
	\centering
	\includegraphics[width=0.7\textwidth]{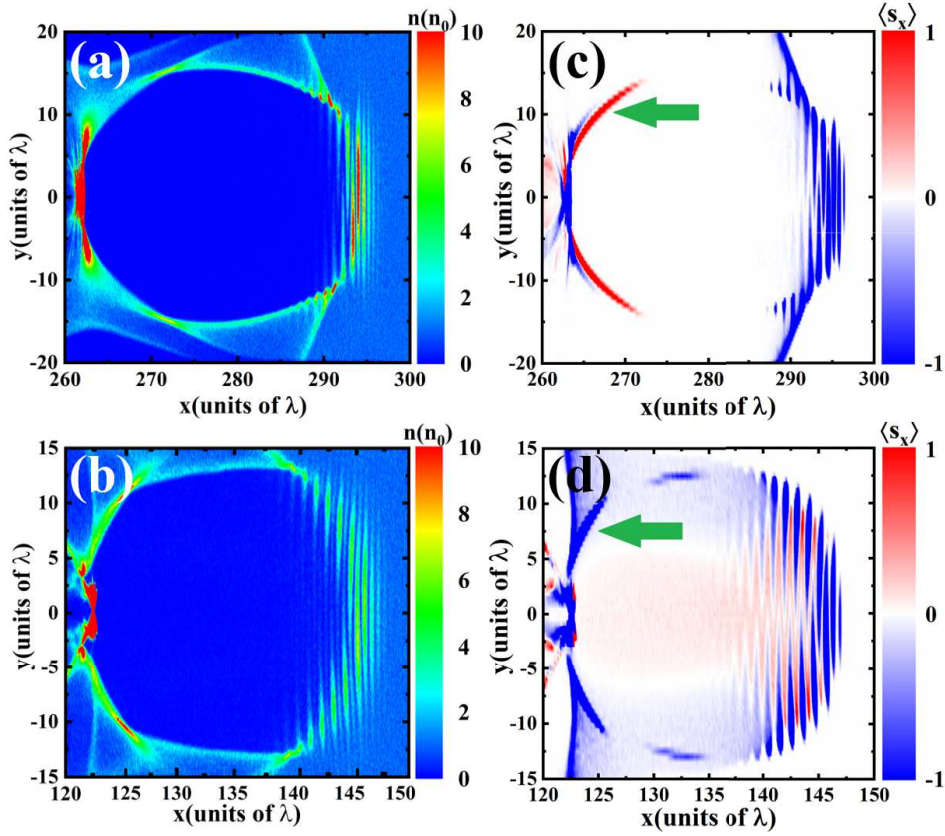}
	\caption{(color online) Electron density snapshots of prolate ($\eta=0.92$) and oblate ($\eta=1.21$) bubbles obtained from 3D PIC simulations with an initial plasma density of $n_0=0.006n_c$ (a) and $n_0=0.015n_c$ (b), respectively. Snapshots  of the  $\langle{s_x}\rangle$ polarization density of electrons with kinetic energy $E_k>1$ MeV, in prolate  (c) and oblate  (d) bubbles. The $\langle{s_x}\rangle$ of shell electrons are indicated by the green arrows. The laser parameters are same as  in Fig. 1.}
	\label{fig_2}
\end{figure*}

The degree of electron spin precession differs for the three cases. The electrons can stay longer in the second stage with increasing $\eta$, which means that the degree of clockwise rotation is largest for case 3 ($\eta>1$), where $s_x$ approaches $-1$ as shown as Fig. 1(b3). During the third stage, the degree of spin precession in the three cases is also different. In case 1 ($\eta<1$) and case 2 ($\eta=1$), ${s}_x$ increases to 1 (its initial value) and decreases afterwards, while ${s}_y$ changes from negative to positive. In case 3 ($\eta>1$), the counter-clockwise spin precession is significantly smaller. Thus, ${s}_x$ increases from $-1$ to a value near $0$ and ${s}_y$ remains negative.

 During the fourth stage $t>t_{\mathrm{\uppercase\expandafter{\romannumeral3}}}$ the  electrons remain in the acceleration phase and oscillate around the laser axis. The $s_x$ and $s_y$ values  oscillate around their values at $t=t_{\mathrm{\uppercase\expandafter{\romannumeral3}}}$, as shown in the inset of Fig. 1(b1)--(b3). Compared to the strong precession during the earlier stages, this spin variation can be ignored. This means that the spin  procession for electrons \textit{mainly occurs during self-injection}, and does not change significantly thereafter in the acceleration stage, and this result is  consistent with the study of Wen \emph{et al.} \cite{wen2019polarized}.  When the electron arrives at the centre of bubble, the acceleration process is terminated and the spin direction is similar to its value at $t=t_{\mathrm{\uppercase\expandafter{\romannumeral3}}}$, as revealed as insets of Fig. 1(b1)-(b3). At this time, the final spin orientation is forward in case 1 ($\eta<1$), case 2 ($\eta=1$) and backward in case 3 ($\eta>1$). We can thus conclude that the spin precession is strongly affected by the bubble geometry. Moreover, during the electron injection, the clockwise spin rotation during the second stage is partly balanced by the counter-clockwise rotation of stage three. As a consequence, a particular ellipsoidal bubble shape can be chosen for which the electron spins can be restored to their initial orientation at $t=t_{\mathrm{\uppercase\expandafter{\romannumeral3}}}$ and maintain dynamic stability over the acceleration phase.
 
The net polarization of a particle beam is defined as $P=\sqrt{\langle{s_x}\rangle^2+\langle{s_y}\rangle^2+\langle{s_z} \rangle^2}$, where $\langle{s_i}\rangle$ is the average value in each direction. This definition is a statistical average for an electron bunch. An accelerated electron beam can be mimicked by changing the initial position $y_0$ for a set of test particles. The trapping cross-section has been studied using the same method in a previous study \cite{li2014dependence}.  The distributions of ${s}_x$ and ${s}_y$ as a function of initial position $y_0$ at the end of acceleration stage are displayed in the Fig. 1(c1)--1(c3). The electrons are also distinguished according to their final energy, denoted by the colour scale. In 3D geometry, the injected electrons  ($\gamma>50$) are distributed in a ring. The accelerated electrons (red dots) are distributed from $r_\mathrm{min}$ to $r_\mathrm{max}$ in these three cases, which means that the electron charge is  affected by the bubble geometry. More importantly, the spin direction of the accelerated electrons also depends on the bubble shape. The value of $\langle{s_x}\rangle$ of the electron beam can be calculated as, 
 \begin{equation}
 \langle{s_x}\rangle=\frac{\sum_{i}^{N} s_{xi}}{N}=\frac{\sum_{r_{min}}^{r_{max}} s_{xi}{\cdot}2{\pi}y_0\delta{y_0}}{\sum_{r_{min}}^{r_{max}} 2{\pi}y_0\delta{y_0}} \ ,
 \end{equation}
 with $\delta{y_0}= 0.01$ in our simulation. Considering the azimuthal symmetry of the bubble field, we have $\langle{s_y}\rangle=\langle{s_z}\rangle=0$ and $P=\left|\langle{s_x}\rangle\right|$. The polarization amounts to $P=0.55$, 0.70 and 0.09 for cases 1 ($\eta<1$), 2 ($\eta=1$) and 3 ($\eta>1$) from Fig. 1, respectively.
  
 \begin{figure*}[t]
 	\centering
 	\includegraphics[width=0.97\textwidth]{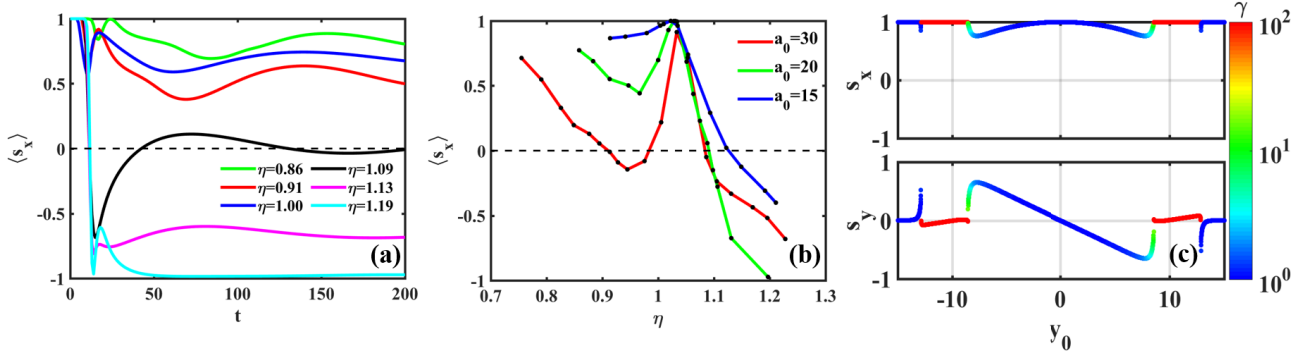}
 	\caption{(color online) (a) Evolution of $\langle{s_x}\rangle$ for an accelerated electron beam in the bubble regime with different shapes ($\eta$). (b)  Dependence of $\langle{s_x}\rangle$ at the end of the acceleration phase as a function of bubble geometry ($\eta$) with different laser intensity. The other laser parameters are same as in Fig. 2. (c)  ${s}_x$ and ${s}_y$ as a function of $y_0$ at the end of acceleration stage for $n_0=0.020n_c$, $\eta=1.04$ and a $a_0=30$ laser pulse. The electron energy is indicated by the dot colors.}
 	\label{fig_3}
 \end{figure*}
 
  \begin{figure*}[t]
 	\centering
 	\includegraphics[width=0.97\textwidth]{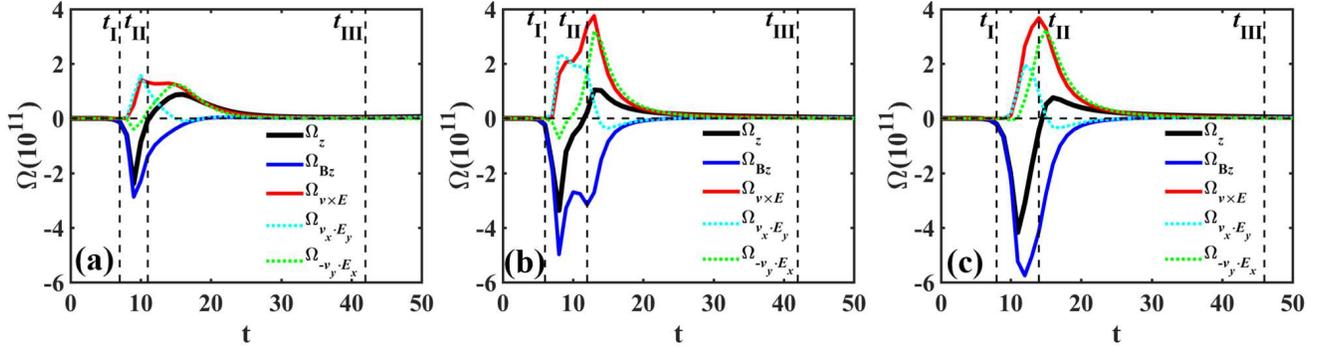}
 	\caption{(color online)  Evolution  of the total spin precession frequency $\bm{\Omega}_T$ (black line), term $\bm{\Omega}_B$ (blue line), term $\bm{\Omega}_{E}$ (red line), term $\bm{\Omega}_{v_x\cdot E_y}$ of $\bm{\Omega}_{E}$ caused by ${v}_x\cdot{E}_y$ (lightblue dashed), and term $\bm{\Omega}_{-v_y\cdot E_x}$ of $\bm{\Omega}_{ E}$ caused by $-{v}_y\cdot{E}_x$ (sea-green dashed) for a typical electron in the bubble with different geometry (a) $\eta=0.91$ (Case 1 in Fig. 1), (b)$\eta=1.04$ (Case in Fig. 3(c)) and (c) $\eta=1.09$ (Case 3 in Fig. 1), respectively.}
 	\label{fig_4}
 \end{figure*}
 
 \section{\label{sec:picsims} 3D PIC SIMULATIONS \& DISCUSSION}
To verify these findings we present results of 3D PIC simulations with a modified version of the EPOCH code \cite{arber2015contemporary}, in which the TBMT equation for the study of electron and ion spins was implemented \cite{li2021polarized}. The laser parameters are the same as in Fig. 1. The initial plasma density is $n_0=0.006 n_c$ and $n_0=0.015  n_c$, respectively.  The size of the moving window is $60\lambda(x)\times40\lambda(y)\times40\lambda(z)$ with resolutions $\mathrm{d}x=\lambda/32$ and $\mathrm{d}y=\mathrm{d}z=5\mathrm{d}x$, and  8 pseudo-particles per cell. 

As shown as in Fig. 2(a) and (b), the simulated bubble shapes are prolate  ($n_0=0.006n_c$) and oblate  ($n_0=0.015n_c$), respectively.
The distribution of $\left\langle s_x \right\rangle$ for the two cases are presented in Fig. 2(c) and (d). The polarization of electrons, located at the tail of bubble, is positive (green arrows) for the prolate  bubble ($\eta=0.92$), while it is negative for the oblate case ($\eta=1.21$). Thus, the results of our 3D PIC simulations verify the results of the test-particle simulations. Here, the electrons located at the shell of the bubble were analysed because of depolarization process mainly occurs before the electron arrives at the rear wall of the bubble. Moreover, the injection process and the evolution of bubble are also affected by the laser, effects which are not included in the earlier test-particle analysis. 
To further illustrate how the electron-beam polarization variation depends on the bubble geometry, we did a series of 2.5D PIC simulations with initial plasma densities ranging from $n_0=0.005n_c$ to $n_0=0.025n_c$ and fixed laser intensity $a_0=20$. Other parameters were the same as in Fig. 1. The data of bubble geometry were substituted in the single electron dynamic simulations. The evolution curves of $\langle{s_x}\rangle$ with time are shown in Fig. 3(a).  Here, bubbles with six typical aspect ratios ($\eta$) were selected. We found that the $\langle{s_x}\rangle$ of the electron bunch first decreases then increases in every case, which is similar to the spin dynamics of a single electron -- Fig. 1.  When $\eta$ increases from $0.86$ to $0.91$, $\langle{s_x}\rangle$ decreases.  When $\eta$>1, $\langle{s_x}\rangle$ also decreases with increasing $\eta$ and even goes negative in an oblate spheroidal bubble. Its  value is maximized in a spherical bubble ($\eta=1$) compared with the two cases with $\eta=0.91$ and $\eta=1.09$. This means that the polarization of an accelerated electron bunch can be preserved when the bubble shape is nearly spherical. 
 
In order to check the persistence of this phenomenon, further 2.5D PIC simulations were carried out for different laser parameters. The dependence of $\langle{s_x}\rangle$ on the various aspect ratios $\eta$ after acceleration with laser intensities $a_ 0=15,20,30$ are presented in Fig. 3(b). It is found that when the bubble is nearly spherical, the electron-beam depolarization is always minimal. It can even be zero as shown in Fig. 3(c), where $n_0=0.20n_c$, $a_0=30$ and $\eta=1.04$, respectively. 
In the case of $a_0\ge20$, the value of $\langle{s_x}\rangle$ is negatively correlated with $\eta$ when $\eta<0.9$, which arises from the smaller number of electrons injected into the bubble.   
 
Finally, the mechanisms governing the spin dynamics in the bubble fields are considered. In Fig. 4 the evolution of the rotation frequency is analysed for  three cases: (a) $\eta=0.91$ (case 1 in Fig. 1), (b) $\eta=1.04$ (case in Fig. 3(c)), where $\langle{s_x}\rangle$=0.99; (c) $\eta=1.09$ (case 3 in Fig. 1).  The terms of Eq. (3) are separated into $\bm{\Omega}_E$ and $\bm{\Omega}_B$ for studying their individual contributions to the rotation frequency. Note that $|\bm{\Omega}_a| \ll |\bm{\Omega}_T|$, and that $\bm{\Omega}_T$ (solid black line) can be divided to $\bm{\Omega}_B={e\bm{B}_z}/{m_e\gamma}$ (solid blue line) and $\bm{\Omega}_{E}=-{e(\bm{v}\times\bm{E})/[{m_e(\gamma+1)c^2}]}$ (solid red line).
 
 During the second stage ($t_{\mathrm{\uppercase\expandafter{\romannumeral1}}}<t<t_{\mathrm{\uppercase\expandafter{\romannumeral2}}}$), the electrons located near ($\xi\approx0$, $y\approx{r_{\perp}}$) mainly feel $E_y$ and $B_z$  as given in Eqs.(2). Then, $\bm{\Omega}_{B_z}$ and the  ${v_x\cdot E_y}$ term in $\bm{\Omega}_{E}$ make the largest  contribution to $\Omega_T$. Since the electrons cannot obtain enough energy, this results in ${ \left| \Omega_{B_z}\right|}>{\left|\Omega_{E}\right|}$ and the spins rotate clockwise. Assuming a bubble velocity $v_\mathrm{b}=1$, we obtain $B_z{\propto}({\eta^2}y)/4$ and $E_y{\propto}[(2-{\eta^2})y]/4$, based on Eqs. (2). With increasing of $\eta$, $B_z$ increases and $E_y$ decreases, then $\Omega_T$ increases. Meanwhile, the electrons stay longer in this phase, which allows the  contribution of $B_z$ to dominate with increasing $\eta$, and the degree of clockwise rotation for electron spin is positively correlated with $\eta$. 
   
 During the third stage ($t_{\mathrm{\uppercase\expandafter{\romannumeral2}}}<t<t_{\mathrm{\uppercase\expandafter{\romannumeral3}}}$), electrons arrive at the tail of bubble ($\xi\approx{R_{\perp}}$, $y\approx{0}$), where $E_y{\approx}0$ and $B_z{\approx}0$. Here the electrons mainly feel $E_x$, then the part  of ${v_y\cdot E_x}$ in $\bm{\Omega}_{E}$ becomes the dominant contribution, which results in counter-clockwise spin rotation. At the tail of bubble $E_x\approx{\eta^2\xi}/2$ under the assumption $v_\mathrm{b}=1$. Considering $R_{\perp}$ is similar for bubbles with different shapes and $\xi=R_{\|}$, we obtain $E_x\approx{R_{\perp}^2}/(2R_{\|}$). With increasing $\eta$, $R_{\|}$ decreases and the part of ${v_y\cdot E_x}$ in $\Omega_E$ increases. Moreover, considering that the times when electron reach the tail of bubble are different, then the contribution of $\Omega_{B_z}$ is  different for these three cases, which results in similar values of $\bm{\Omega}_T$.
 
 For the overall process, the spin rotation is the sum of the clockwise rotation during the second stage and the counter-clockwise rotation during the third stage. With increasing $\eta$, the degree of clockwise rotation increases and the degree of counter-clockwise rotation stays roughly constant.  Clockwise rotation dominates in the prolate bubble, whereas the counter-clockwise rotation is prevalent in an oblate bubble. More importantly, the two  precessions can cancel each other in a bubble with a certain value of $\eta$. As shown as Fig. 4(b), an accelerated electron beam with no net depolarization can be produced when $\eta=1.04$.
 
 \section{\label{sec:level3}CONCLUSION}
In summary, the depolarization of accelerated electron bunches has been studied through test-particle dynamics during the interaction of a high-intensity laser with a longitudinally pre-polarized plasma. Spin rotation occurs mainly during the self-injection process. As a consequence, the bubble geometry has a strong influence during this phase, since it determines the electromagnetic field distributions. It is also found that  polarization is preserved for nearly spherical bubble shapes. In contrast, non-spherical bubble shapes lead to strong depolarization. These findings should help to choose suitable laser-plasma parameters for producing a polarized electron beam using a pre-polarized plasma. Moreover, the case of a transversal polarization plasma, which is better accessible experimentally, will be investigated in the near future.

\begin{acknowledgments}
The 3D PIC simulations were carried out on the JURECA supercomputer at J\"{u}lich Supercomputing Centre, in particular through the computing time projects JZAM04 and LAPIPE. This work was supported by Germany Postdoctoral Council and the Helmholtz Centre (Grant No. 20191016) and China Postdoctoral Science Foundation (Grant No.2018M641993). The work of M.B. was carried out in the framework of the J\"ulich Short-Pulse Particle and Radiation Center \cite{buscher2020jusparc} and was supported
by the Accelerator Technology Helmholtz Infrastructure consortium ATHENA.  This work was also supported by the Strategic Priority Research Program of Chinese Academy of Sciences (Grant No.XDA25050100), the National Natural Science Foundation of China (No.11804348, No.11775056, No.11975154, and No.11991074), and the Science Challenge Project (No.TZ2018005).
\end{acknowledgments}



\end{document}